\documentclass{jfm}
\usepackage{lipsum}
\usepackage{multicol}
\usepackage{floatrow}
\newfloatcommand{capbtabbox}{table}[][\FBwidth]

\usepackage{graphicx}
\usepackage{newtxtext}
\usepackage{newtxmath}
\usepackage{natbib}
\usepackage{hyperref}
\hypersetup{
    colorlinks = true,
    urlcolor   = blue,
    citecolor  = black,
}

\newcommand{\RomanNumeralCaps}[1]

\title{Multi-scale interactions in turbulent mixed convection drive efficient transport of Lagrangian particles}

\author{Andrew P. Grace\aff{1}
  \corresp{\email{agrace4@nd.edu}} \and
  David Richter\aff{1}
}

\affiliation{\aff{1}Department of Civil and Environmental Engineering and Earth Sciences, University of Notre Dame,
Notre Dame, Indiana 46556, U.S.A.}

\begin{document}

\maketitle

\begin{abstract}
When turbulent convection interacts with a turbulent shear flow, the cores of convective cells become aligned with the mean current, and these cells (which span the height of the domain) may interact with motions closer to the solid boundary. In this work, we use coupled Eulerian-Lagrangian direct numerical simulations of a turbulent channel flow to demonstrate that under conditions of turbulent mixed convection, interactions between motions associated with ejections and low-speed streaks near the solid boundary, and coherent superstructures in the interior of the flow interact and lead to significant vertical transport of strongly settling Lagrangian particles. We show that the primary suspension mechanism is associated with strong ejection events (canonical low-speed streaks and hairpin vortices characterized by $u'<0$ and $w'>0$), whereas secondary suspension is strongly associated with large scale plume structures aligned with the mean shear (characterized by $w'>0$ and $\theta'>0$). This coupling, which is absent in the limiting cases (pure channel flow or free convection) is shown to lead to a sudden increase in the interior concentration profiles as $\mathrm{Ri}_\tau$ increases, resulting in concentrations that are larger by roughly an order of magnitude at the channel midplane.
\end{abstract}


\section{Introduction}


It is well-known through numerical simulation, experiment, and theory that fluid turbulence significantly influences the transport of heavy Lagrangian particles with appreciable inertia. For example, turbulence generated near Earth's surface can lead to long range transport ($>2000$ km) of giant dust grains ($>100 \mu$m) \citep{adebiyi_review_2023,van_der_does_mysterious_2018}, while simple scaling theories predict a much shorter displacement due to their strong settling behaviour. Understanding and predicting the ultimate fate of heavy particles is particularly interesting to a wide range of scientific disciplines due to the climatic and health impacts of atmospheric aerosols and particles, and accurate modeling of the primary mechanisms responsible for this transport are of key interest. In this ``mysterious long-range transport" problem,
\citet{van_der_does_mysterious_2018} hypothesized several mechanisms which may influence particle transport including strong convection and high wind speeds, on which we place our primary focus. Interestingly, under convective conditions and moderate to strong mean shear, coherent, roll-like structures arise, and are well documented in Earth's atmosphere \citep[and references therein]{kuo_perturbations_1963,salesky_nature_2017}. In idealized studies (discussed more below), this phenomenon is often referred to as ``mixed convection". In this work, our goal is to investigate the consequences of mixed convection on the suspension and transport of strongly settling inertial particles.

When turbulence production by both convection and shear are comparable, the flow is termed ``mixed convection" (the limiting cases being a pure channel flow and free convection), and it is this dynamic regime in which we focus our study.
Though our knowledge of free and forced convection is extensive (see \cite{grossmann_scaling_2000,lohse_ultimate_2024}, for example),
the mixed convection literature is comparatively small. For example, the first study focused on the role of shear in Rayleigh-B\'{e}nard (RB) convection was undertaken by \cite{domaradzki_direct_1988}, and since then, there have been numerous studies focused on the impacts of mixed convection of the vertical transport of heat in channel flows \citep{scagliarini_heat-flux_2014,scagliarini_law_2015,pirozzoli_mixed_2017}, and atmospheric boundary layers \citep{moeng_comparison_1994,salesky_nature_2017}. A primary hallmark of mixed convection is that convective plumes generated near a solid boundary become aligned with the background flow within the interior of the flow, leading to large, coherent streamwise rollers, often referred to as superstructures.  

Studies focused on the role of convection on the transport of inertial particles is relatively few. For example, see the recent work by \citet{denzel_stochastic_2023} who focused on developing a stochastic model for the lifetime of small particles in an experimental chamber designed to study clouds in RB flow \citep{chang_laboratory_2016}. To the authors' knowledge, there may only be one study focused on the role of mixed convection in particle transport \citep{zaza_mixed_2024}. In that work, the authors focused on the role of the particles in heat transfer throughout a turbulent boundary layer, and while the authors traversed a relatively large range of Richardson number (the key parameter quantifying the strength of convective turbulence to shear generated turbulence), they were restricted to relatively low Reynolds number ($\mathrm{Re_\tau}=180$), and ignored particle settling.
In this work, we aim to describe a new mechanism by which strongly settling inertial particles are efficiently mixed through the boundary layer, in a way that isn't present for either pure channel flow or free convection. Specifically, we are interested in the interactions between the near-boundary structures (associated with low-speed streaks and ejection events, see e.g., \citet {wallace_quadrant_2016}), and the interior superstructures generated by convective plumes, and how they couple to lead to vertical transport of isothermal inertial particles settling under the action of gravity. To investigate the dynamics, we use a series of coupled Eulerian-Lagrangian direct numerical simulations to simulate settling inertial particles channel flow ranging from free convection to pure shear. 

\section{Technical Background \label{sect:background}}
\subsection{Carrier Phase \label{subsect:carrier_phase}}

In this letter, we use the NCAR Turbulence with Lagrangian Particles Model \citep{richter_inertial_2018} to simulate one-way coupled inertial particles emitted from the lower solid boundary a turbulent closed-channel flow. This code has been validated and used in multiple studies focused on inertial particle settling and transport in turbulent boundary layers \citep{wang_inertial_2019,bragg_mechanisms_2021,gao_direct_2023,grace_reinterpretation_2024}. For the carrier phase, we use direct numerical simulations (DNS) to solve the three-dimensional, incompressible Navier-Stokes equations under the Boussinesq approximation in a turbulent channel flow setup of streamwise length $L_x$, spanwise extent $L_y$, and total height $2h$. At the upper and lower boundaries, a no-slip boundary condition is enforced, while the domain is periodic in the $x$ and $y$ directions. The background state of the carrier phase is established by accelerating the flow with an imposed pressure gradient, $-dP/dx>0$ (note that $\hat{\boldsymbol{x}}$ is the unit vector in the streamwise direction) and allowing the flow to become turbulent. The magnitude of the pressure gradient allows us to define a friction velocity $u_\tau = \sqrt{\tau_w/\rho_a}$, where $\tau_w$ is the stress at the lower boundary and $\rho_a$ is the fluid density. 
The governing parameters of the carrier phase are:
\begin{equation}
    \mathrm{Ri}_\tau = \frac{\mathrm{Ra}}{\mathrm{PrRe}_\tau^2},\quad \mathrm{Ra} = \frac{g\alpha_\theta\Delta T(2h)^3}{\nu\kappa}, \quad \mathrm{Re}_\tau = \frac{u_\tau h}{\nu},\quad \mathrm{Pr} = \frac{\nu}{\kappa}. 
\end{equation}
Respectively, these are the Richardson number, the Rayleigh number, the friction Reynolds number, and the Prandtl number. In these parameters, $\alpha_\theta$ is the isobaric thermal expansion coefficient. Throughout this work, we assume $\mathrm{Pr}=0.715$ for all cases, where $\nu$ is the kinematic viscosity for dry air, and $\kappa$ is the thermal diffusivity of dry air. 
As mentioned previously, $\mathrm{Ri_\tau}$ is an important parameter as it characterizes the relative importance of turbulence generated by convection (through $\mathrm{Ra})$ to that generated by friction along the solid boundary (through $\mathrm{Re}_\tau)$. Practically speaking, \citet{pirozzoli_mixed_2017} noted the appearance of coherent superstructures in both the temperature field and vertical fluctuating velocity field within the regime $1< \mathrm{Ri}_\tau < 1000$. In this work, we aim to investigate the regime $\mathrm{Ri_\tau}\sim \mathcal{O}(10-100)$ as it allows us to attain computationally feasible values for $\mathrm{Re_\tau}$ while also allowing for the formation of the coherent roll structures. 
Values for $\mathrm{Ra}$, $\mathrm{Re}_\tau$, and $\mathrm{Ri}_\tau$ used throughout this work can be found in Table \ref{tab:cases}. In the table, case names correspond to the dynamic regime of the flow, i.e. CF (channel flow), FC (free convection), and MC (mixed convection). Note there is a case named MC-Sc, which corresponds to a mixed convection case with a lower Schmidt number value (discussed next).

\subsection{Dispersed Phase}
The applications of this work are towards coarse dust transport in the atmospheric surface layer. These dust particles range in size, but even coarse and giant grains (roughly 30-100 $\mathrm{\mu m}$) are significantly smaller than the local Kolmogorov scale, which can be in the range of several millimetres. These particles are also significantly denser than the carrier phase, and their volume fractions are low once they are above the emission layer, so we may ignore added mass and Basset-History forces, as well as two-way coupling and particle-particle interactions. Given these assumptions, we apply the point-particle approximation and apply the conservation of momentum for a rigid spherical particle subjected to linear hydrodynamic drag and gravity. The one-way coupled point particle approach also has the added benefit that each particle is independent from each other particle, effectively removing the volume fraction as a governing parameter. This allows us to increase the number of particles to ensure convergence of the statistics of interest without affecting the flow. 

Using the local Kolomogorov scales to non-dimensionalize the particle equations of motion, we arrive at 
\begin{equation}
    \mathrm{St}_\eta\frac{d\boldsymbol{v}_p}{dt} =\Psi\left(\boldsymbol{u}_f(\boldsymbol{x}_p(t),t)- \boldsymbol{v}_p\right) - \mathrm{Sv}_\eta\hat{\boldsymbol{z}} 
\end{equation}
Here, $\boldsymbol{v}_p = (v_1,v_2,v_3)$ is the three dimensional velocity vector for each particle, $\boldsymbol{x}_p$ is the location of each particle in space, $\boldsymbol{u}_f(\boldsymbol{x}_p(t),t) = (u,v,w)$ is the three dimensional instantaneous flow velocity evaluated at the location of the particle. $\mathrm{St}_\eta$ and $\mathrm{Sv}_\eta$ are the governing parameters of the particle equations of motion written in terms of the local Kolmogorov scales, which can be defined in channel flow, mixed convection, and free convection. We also report the governing parameters in terms of the viscous scales of the flow, $\mathrm{St}^+$ and $\mathrm{Sv}^+$, which are defined in channel flow and mixed convection, and $\mathrm{St}_*$ and $\mathrm{Sv}_*$ which are defined in mixed and free convection. These parameters are defined as
\begin{equation}
    \mathrm{St}_\eta =\frac{\tau_p\epsilon^{1/2}}{\nu^{1/2}}, \;\mathrm{Sv}_\eta =\frac{v_g}{\left(\nu\epsilon\right)^{1/4}},\; \mathrm{St}^+ =\frac{\tau_pu_\tau^2}{\nu}, \; \mathrm{Sv}^+ =\frac{v_g}{u_\tau},\; \mathrm{St}_* =\frac{\tau_pw_*}{h}, \; \mathrm{Sv}_* =\frac{v_g}{w_*},
\end{equation}
and their values used throughout this work can be found in table \ref{tab:cases}. In the expressions above, $\epsilon$ is the domain averaged dissipation, $w_* = (g\alpha h \langle w'\theta'\rangle_s)^{1/3}$ is the Deardorf convective velocity scale ($\langle w'\theta'\rangle_s$ is the surface heat flux), and $\tau_p$ is the Stokes relaxation timescale of the particles. 
The Stokes settling velocity is defined as $v_g = \tau_p g_p$ where $g_p$ is the gravitational acceleration applied to the particle. In simulations, $g_p$ need not be equivalent to $g$ (the gravitational acceleration applied to the fluid), thus allowing for turbulence, settling, and inertial properties to be specified independently. $\Psi = 1 + 0.15\mathrm{Re}_p^{0.687}$ is the Schiller-Neumann correction to the drag force, and $\mathrm{Re}_p$ is the particle Reynolds number. As we keep the particle diameter fixed for all cases in this work, the particle Reynolds number remains small, meaning that $\Psi\approx 1$.

Next, we must consider is the boundary conditions for the particles, specifically the scheme by which we are to lift the particles into the domain. Here we take an approach focused on simplicity; following \citet{richter_inertial_2018} and \citet{bragg_mechanisms_2021}, we add a Brownian-like jump term to the particle equations of motion. Using this term, particles take a discontinuous jump with zero mean and unit variance $\sqrt{2\kappa_p dt}$ where, for practical purposes, $dt$ is the timestep of the model and $\kappa_p$ is a parameter modeling the particle diffusivity. Mathematically, the location of the particle centroid is advanced according to the following equation:
\begin{equation}
    d\boldsymbol{x}_p = \boldsymbol{v}_p dt + \left(\frac{\mathrm{Sc}}{2}\right)^{-1/2} dt^{1/2}d\boldsymbol{\xi},
\end{equation}
where $d\boldsymbol{\xi}$ is a Weiner process.
Particles are initialized in a thin reservoir of thickness $D$ held at a fixed concentration beneath the lower solid boundary and are emitted through the lower surface. This approach effectively creates a Dirichlet condition for the particle concentration which is maintained throughout the simulation. For the upper boundary, particles reflect creating a Neumann no-flux condition.
This model introduces several new parameters, specifically the particle Schmidt number, $\mathrm{Sc} = \nu/\kappa_p$ and the fixed reservoir concentration, $\mathcal{C}$.

\begin{figure}[H]
    \centering
    \includegraphics[width=0.8\textwidth]{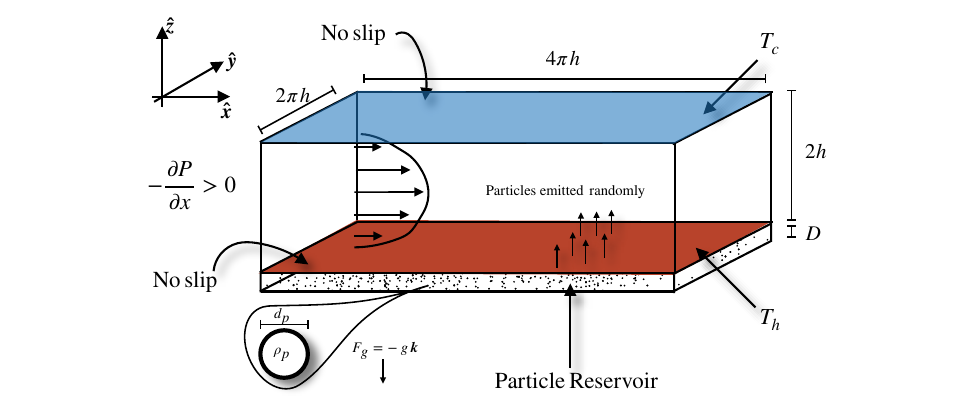}
    \caption{A rectangular channel of dimensions  $4\pi h\times 2\pi h\times 2h$. The flow is periodic in the horizontal and is driven by a constant streamwise pressure gradient. No-slip boundary conditions are enforced at the top and bottom boundaries. The solid boundaries are held at a temperature difference $\Delta T=T_h-T_c$. Particles are emitted randomly from a reservoir (fixed $\mathcal{C}\approx 4000$) into the domain through the bottom boundary. Particles reflect elastically off the upper boundary.}
    \label{fig:schematic}
\end{figure}

A schematic highlighting the salient features of the model is shown in figure \ref{fig:schematic}. For a more realistic (but significantly more complex) approach to particle emission, see \cite{dupont_modeling_2013}, for example.

\begin{table}
    \begin{center}
    \begin{tabular}{lcccccccccccc}                            
        Case name & $\mathrm{Re}_\tau$ & $\mathrm{Ra}$ & $\mathrm{Ri}_\tau$ & $\mathrm{Sc}$ & $\mathrm{St}^+$  & $\mathrm{Sv}^+$ & $\mathrm{St}_\eta$ & $\mathrm{Sv}_\eta$ & $\mathrm{St}_*$ & $\mathrm{Sv}_*$   \\ \hline \hline
        CF                  & 496                       & 0                         & 0         & 0.19  & 6.21  & 0.53  & 0.94  & 1.36  & --     & --   \\
        MC                  & 519                       & $8.4\times 10^{6}$        & 43        & 0.19  & 6.82  & 0.5   & 1.01  & 1.31  & 0.01   & 0.95 \\
        MC$-\mathrm{Sc}$    & 501                       & $8.4\times 10^{6}$        & 46         & 1     & 6.36  & 0.52  & 0.90  & 1.39  & 0.01   & 0.95 \\
        FC                  & --                        & $8.4\times 10^{6}$        & $\infty$  & 0.19  & --    & --    & 0.12  & 3.85  & 0.01   & 1.03 \\
    \end{tabular}
    \caption{Cases discussed in this manuscript. CF stands for ``Channel Flow", MC stands for ``Mixed Convection", and FC stands for ``Free convection". Parameter definitions can be found in the main text. All cases were run on a $512^3$ grid, with $\mathrm{Pr} = 0.715$ and a reservoir concentration of $C\approx 4000$.}
    \label{tab:cases}
    \end{center}
\end{table}

\section{Results\label{sect:results}}
\begin{figure}
    \centering
    \includegraphics[width=\textwidth]{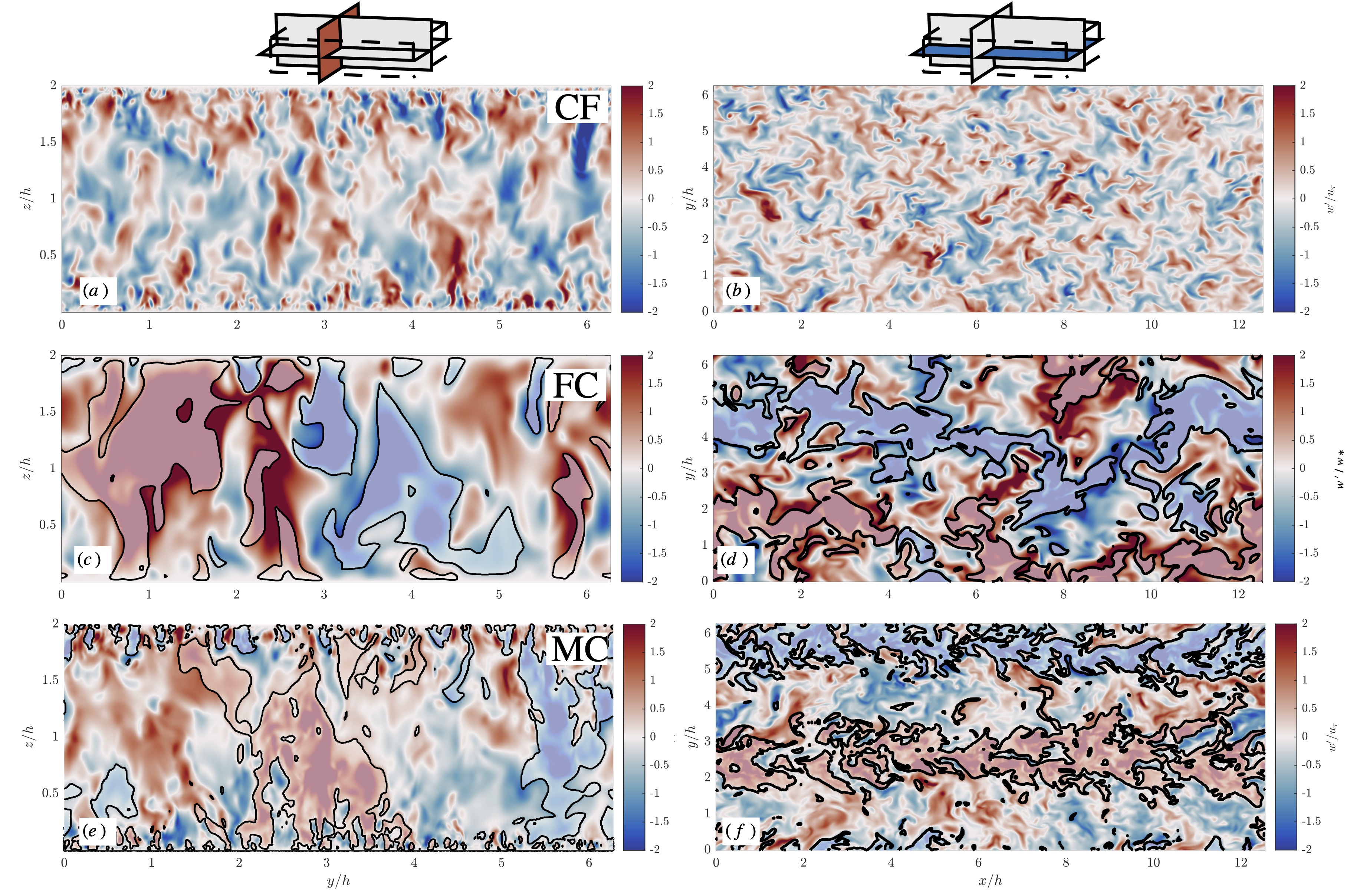}
    \caption{Slices of the fluctuating vertical velocity at the mid-plane for pure channel flow (a-b), mixed convection (c-d), and free convection (e-f). Fluctuating velocities are normalized by $u_\tau$ in panels (a-d) and by $w_*$ (the convective velocity scale) in panels (e-f). The left column shows $x-y$ slices and the right column shows $y-z$ slices. Shaded contours in panels (c-f) show regions of $ w'\theta' > 0.12\kappa\Delta T(2h)^{-1}\mathrm{Ra}^{1/3}$.}
    \label{fig:wslices}
\end{figure}

To examine the consequences of mixed convection on the suspension of settling particles, we first provide a comparison of the coherent structures present in the turbulent flow by examining the fluctuating vertical fluid velocity for a pure turbulent channel flow (CF; figure \ref{fig:wslices} (a-b)), free convection (FC; figure \ref{fig:wslices} (c-d)), and mixed convection (MC; figure \ref{fig:wslices} (e-f)). For each of these cases, the left column shows $y-z$ slices along the centre of the domain, while the right column shows $x-y$ slices at the mid-plane, indicated by small schematics at the top of the figure columns. Finally, for the mixed and free convection cases, we have indicated regions where $w'\theta' > 0.12\kappa\Delta T(2h)^{-1}\mathrm{Ra}^{1/3}$ with contour shading. This criterion is derived from the scaling relationship discussed in \citet{pirozzoli_mixed_2017}, who studied turbulent channel flows under unstable stratification for similar $\mathrm{Re}_\tau$ and $\mathrm{Ra}$. This criterion serves as a quantitative indicator of large positive turbulent heat fluxes in the domain. By isolating regions where the turbulent heat flux is larger than this value (the shaded regions), we can identify structures in the flow that exhibit very strong updrafts and downdrafts. 

The CF case, figure \ref{fig:wslices}(a-b), exhibits characteristic flow structures that vary in vertical scale across the domain, including small scale instabilities associated with low speed streaks near the solid boundaries, and larger features within the interior that scale with $h$. Conversely, in the FC case, shown in \ref{fig:wslices}(c-d), we can see evidence of organized convective cells in both the horizontal and vertical slices, which tend to scale with the full domain height $2h$ and dominate the interior motion. Importantly, there is much less activity near the solid boundary, as the boundary stresses induced by the convective plumes are not significant at this $\mathrm{Ra}$ \citep{pirozzoli_mixed_2017,blass_flow_2020}. The important insight is that MC, figure \ref{fig:wslices}(e-f), shares aspects of both the CF and FC cases. For example, we can see large scale structures within the interior of the MC case (strong updrafts and downdrafts), figure \ref{fig:wslices}(e), reminiscent of the domain size convective cells from the FC case. However, in the horizontal, figure \ref{fig:wslices}(f), these interior plumes becomes strongly aligned in the streamwise direction, with weaker fluctuations between the coherent roll structures. Furthermore, we can see that the near boundary structures (also associated with strong heat fluxes), are qualitatively similar to the CF case. To summarize, MC exhibits large scale convective interior plumes that scale with $2h$, characteristic of FC, but also exhibits significant activity near the solid boundaries, characteristic of CF. However, due to the mean shear, the convective plumes align in the streamwise direction, creating large scale superstructures.  

\begin{figure}
    \centering
    \includegraphics[width=\textwidth]{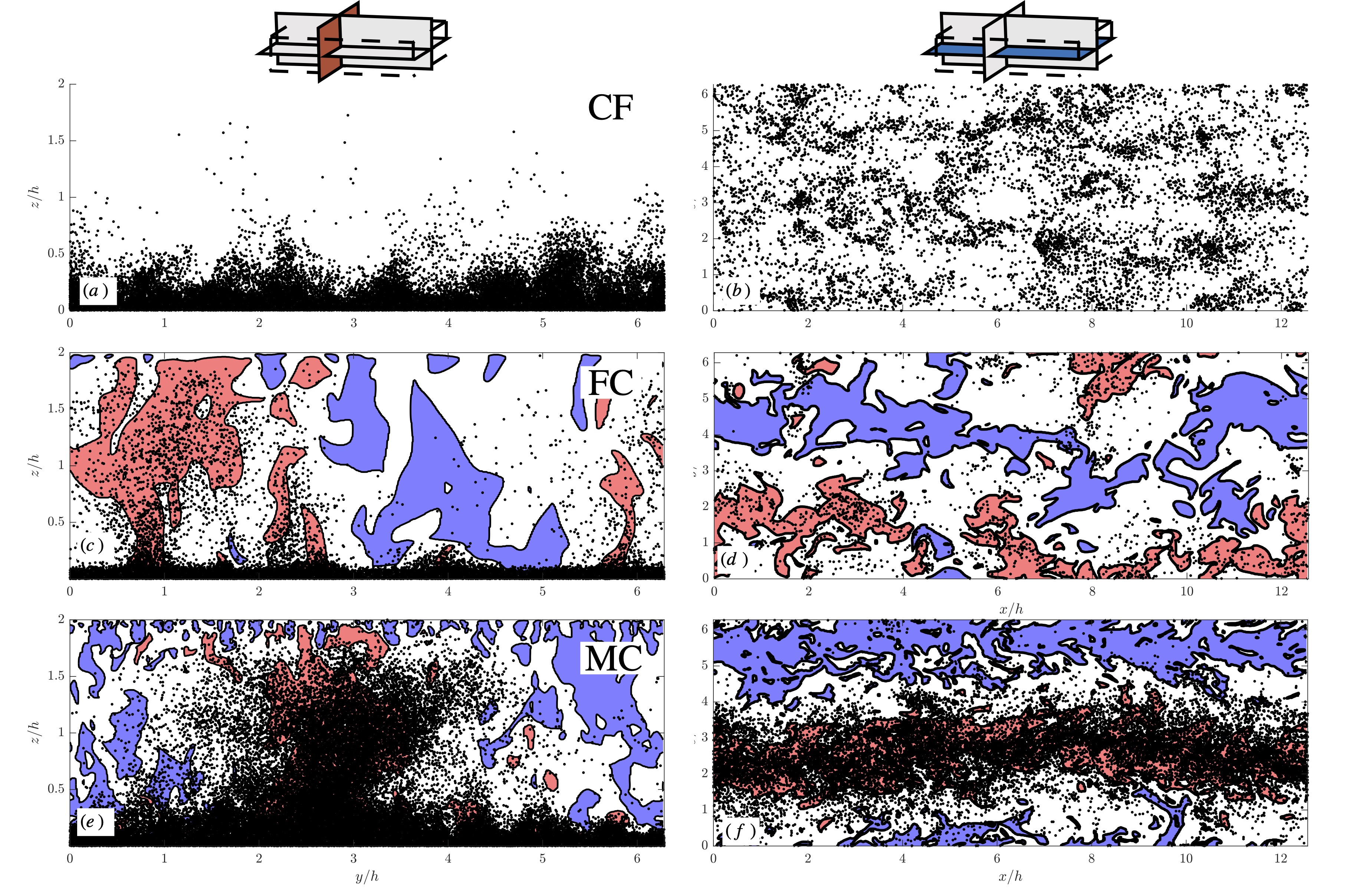}
    \caption{As in figure \ref{fig:wslices}. Contours are regions where $w'\theta' > 0.12\kappa\Delta T(2h)^{-1}\mathrm{Ra}^{1/3}$, and coloured based on the sign of the vertical fluid velocity (red is positive and blue is negative). Particles (not to scale) are overlaid highlighting their clustering behaviour.}
    \label{fig:slices_w_particles}
\end{figure}

We now highlight the role that these structures have on the transport of particles into the interior of the channel in MC, as compared to FC and CF.
Figure \ref{fig:slices_w_particles} shows the snapshots of the high heat flux contours discussed in \ref{fig:wslices}, except here we color the contours based on direction of the vertical velocity fluctuation (red indicates positive fluid velocities while blue indicates negative fluid velocities). Overlaid are particles in a slab of non-dimensional thickness 0.02. It is clear from slices in the CF case, \ref{fig:slices_w_particles}(a), that particles are ejected by the structures near the boundary layer, resulting in some clustering in the mid-plane, figure \ref{fig:slices_w_particles}(b) \citep{lee_effect_2019}. When compared to the FC case, figures \ref{fig:slices_w_particles}(c-d), particles are suspended much higher in the domain when they coincide with a strong updraft, shown in figure \ref{fig:slices_w_particles}(c). However, these updrafts are much less efficient at generating ejections near the boundary, resulting in far fewer particles at the mid-plane, shown in figure \ref{fig:slices_w_particles}(d).  However, since both the near-boundary instabilities and interior convective motions are present in MC, figure \ref{fig:slices_w_particles}(e-f), these structures can interact, and we see a marked increase in the number of particles at the mid-plane in the domain, as well as strong spatial clustering within the longitudinally aligned updraft. This observation suggests a coupling between the canonical near-wall ejections and the large scale convective plumes in the interior, which act cooperatively to lift particles away from the solid wall. In CF and MC, particles are initially ejected via the near-wall instabilities, which we refer to as primary ejection. In the event a primary ejection aligns with a large scale interior plume, particles are entrained into the plume and experience significant vertical transport, termed secondary ejection. Conversely, at these values of $\mathrm{St}^+$ and $\mathrm{Sv}^+$, particles that do not align with a convective plume after primary ejection instead settle back towards the lower boundary. Finally, since the near-boundary instabilities in the FC case are much weaker for this $\mathrm{Ra}$, the mechanism responsible for primary ejection is much weaker (as the interior convective instabilities are much less efficient at removing them from this layer), resulting in far fewer particles suspended in the domain interior. 

\begin{figure}
    \centering
    \includegraphics[width=0.8\textwidth]{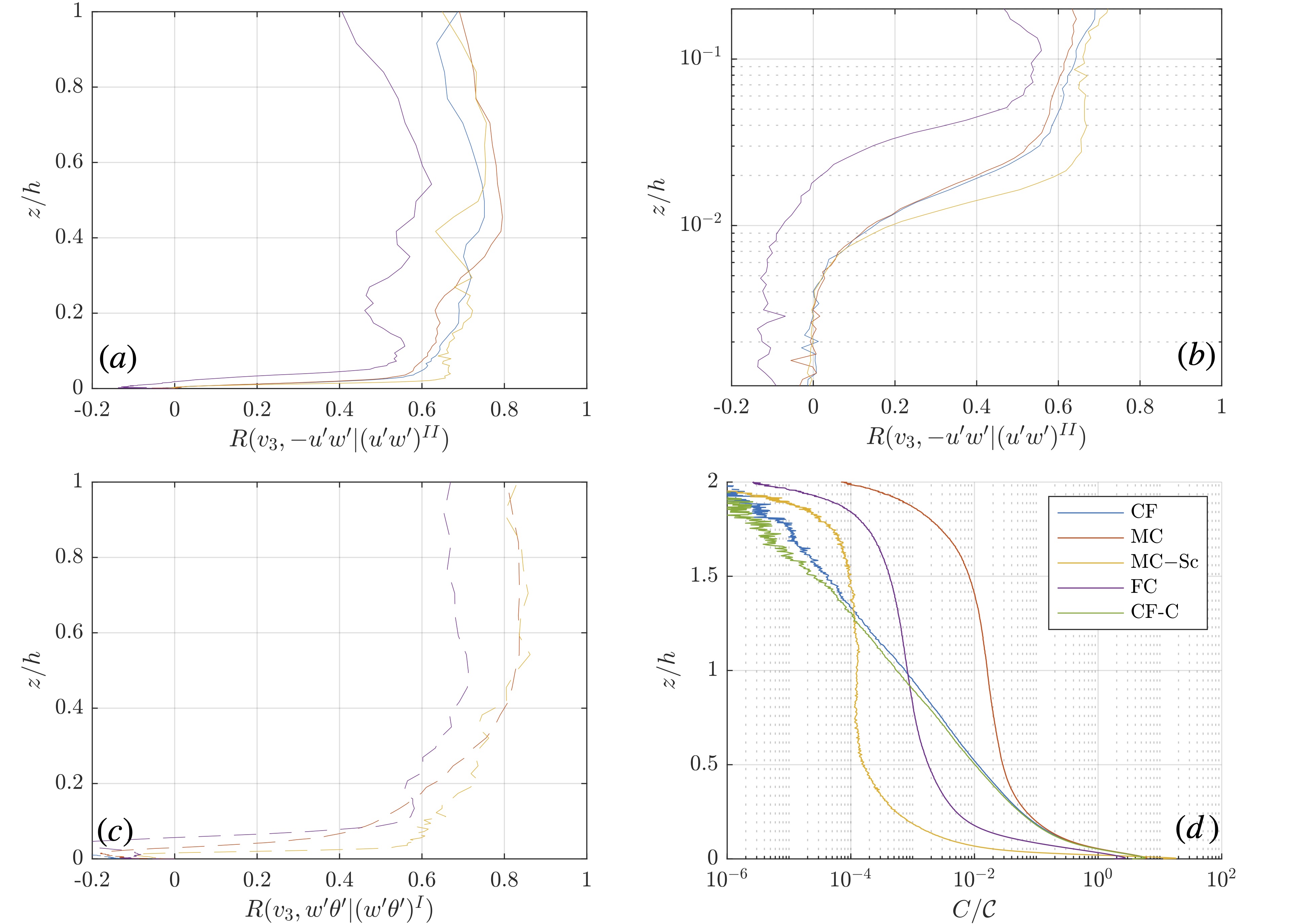}
    \caption{Profiles of slabwise correlation coefficients conditioned on ejection events (a-b), and positive heat fluxes (c). Panel (d) shows concentration profiles averaged in the horizontal dimensions. Also included is a case with $\mathrm{Ri}_\tau=5$ demonstrating the sudden onset of interior mixing as $\mathrm{Ri}_\tau$ increases. }
    \label{fig:correlation_and_concentration}
\end{figure}

Figures \ref{fig:wslices} and \ref{fig:slices_w_particles} serve to qualitatively highlight the multi-scale flow features and associated the particle response within the domain. In figure \ref{fig:correlation_and_concentration}, we use conditional averaging techniques to show slab-wise correlation coefficients for each case discussed in table \ref{tab:cases}. We define the conditional correlation coefficient as 
\begin{equation}
    R(\alpha,\beta|\gamma) = \frac{\langle \alpha\beta|\gamma\rangle}{\sqrt{\langle\alpha^2|\gamma\rangle \langle\beta^2|\gamma\rangle}}, \label{correlation metric}
\end{equation} 
where $\alpha$ and $\beta$ are dummy variables used for demonstration in \eqref{correlation metric}, and $\langle \cdot \rangle$ indicates a slab-wise ensemble average. 
Figure \ref{fig:correlation_and_concentration}(a) and (b) show profiles of the correlation coefficient between the vertical particle velocity and the Reynolds stresses conditioned on ejection events (i.e. $u'<0$ and $w'>0$, where a prime indicates a fluctuating quantity) across the bottom half of the domain, \ref{fig:correlation_and_concentration}(a), and in the bottom 20\% of the domain, \ref{fig:correlation_and_concentration}(b). Following the notation of \citet{salesky_nature_2017}, we use $(u'w')^{II}$ to represent this conditioning. Profiles in figure \ref{fig:correlation_and_concentration}(a) show a moderate to strong correlation between particle velocities and Reynolds stresses in the interior for CF, MC, and MC-Sc, and weaker correlation for FC.  We can see by consulting figure \ref{fig:correlation_and_concentration}(b) that there is rapid increase in correlation near the bottom boundary. This demonstrates that particle velocities are uncorrelated with ejections very near the solid boundary due to the artificial Brownian diffusion (i.e. $z/h<0.01$), followed by a substantial increase in their correlation as they approach primary ejection regions from below. These ejection events manifest themselves as hairpin vortices associated with low-speed streaks, and are present in both CF, MC, and MC-Sc (MC-Sc actually shows a slightly stronger correlation, as expected given the lower artificial diffusion). Moreover, the correlation is markedly lower in FC, as those same low speed streak structures are absent in that case, as discussed before. 

Figure \ref{fig:correlation_and_concentration}(c) shows the particle vertical velocities conditioned on positive heat fluxes (i.e. $w'>0$ and $\theta'>0$). We can see there is a strong correlation in the interior MC and MC-Sc, and a weaker correlation in FC. The correlation is here is likely linked to the spatial coherence of the interior plumes. Moreover, in MC and MC-Sc, these superstructures are typically associated with regions of low horizontal velocity \citep{pirozzoli_mixed_2017}, which explains the strong correlation in the interior, shown in figure \ref{fig:correlation_and_concentration}(a). 

These figures provide quantitative evidence to the hypothesis that primary suspension occurs because of the ejection events associated with hairpin voritices and Reynolds stress, and secondary suspension within the interior occurs as particles are entrained by streamwise oriented convective rolls which are responsible for the bulk heat flux. Importantly, this coupling is only present in mixed convection, and absent in both limiting cases, and has implications for the global profiles of concentration throughout the domain. 
The net effect of the coupling present in MC is that there is a significant and non-monotonic increase in the interior concentration relative to both limiting cases, shown in figure \ref{fig:correlation_and_concentration}(d). While the concentration profiles for MC and CF are coincident below $z/h\approx 0.25$, we see a departure of MC above this height, leading to progressively larger differences in the interior concentration, and even up to and order of magnitude at the mid-plane. Moreover, we have included a case with $\mathrm{Ri}_\tau =5 $, demonstrating that the interior entrainment by secondary suspension remains weak until $\mathrm{Ri_\tau}$ becomes large enough that the interior convective plumes align --- a process that happens very rapidly between $\mathrm{Ri}_{\tau} = 5$ and $\mathrm{Ri}_{\tau} = 43$. Beyond this value of $\mathrm{Ri}_{\tau}$, we note the similarity in shape between MC, MC-Sc, and CF in the interior, even as the absolute concentration changes. These results highlight the cooperative action of primary suspension (by ejections) and secondary suspension (by aligned convective plumes): both working together leads to more efficient distribution throughout the full boundary layer than either one alone.

\section{Conclusions and Discussion\label{sect:conclusions}}
In this work, we have provided the first demonstration of the role of turbulent mixed convection in the suspension of strongly settling Lagrangian particles. Through the use of coupled Eulerian-Lagrangian direct numerical simulations of turbulent mixed convection, achieved here with $\mathrm{Ra}\approx 10^7$ and $\mathrm{Re_\tau}\approx 500$, we provided evidence for a distinct, multi-scale mechanism that can serve as an efficient means of inertial particle suspension (characterized by $\mathrm{Sv}_\eta \sim\mathcal{O}(1)$ and $\mathrm{St}_\eta \sim\mathcal{O}(1)$). We showed that when particles are ejected near the solid boundary (termed primary suspension), they may become entrained within streamwise aligned interior convective plumes (termed secondary suspension), leading to significant vertical transport, and clustering. 


By considering correlation coefficients of vertical particle velocities conditioned on ejections and regions of strong positive heat fluxes, our observations (at $\mathrm{Ri}_\tau \approx 40$) suggest that the action of the near-boundary ejections and the streamwise aligned interior plumes couple together to lead to an efficient suspension mechanism for Lagrangian particles. This occurs despite the large particle settling velocity and inertia, leading to an increase in midplane concentration by roughly an order of magnitude when compared to the limiting cases. Importantly, this cooperative action occurs due to the strong alignment of the convective plumes and is effectively absent in the limiting cases (pure channel flow and free convection), and when convection is present, but weak enough that convective plumes do not become aligned (observed at $\mathrm{Ri}_\tau\approx 5$ here).

These results have important implications for both fundamental and applied studies on the influence of mixed convection on particles, including a potentially strong influence on clustering and dispersion, as well as determining particle residence times. Moreover, it is known that these fluid structures appear ubiquitously at the field scale in the planetary boundary layer \citep{moeng_comparison_1994,salesky_nature_2017}, so more work should take place to understand these cooperative dispersed phase transport mechanisms in the natural environment.

\section{Acknowledgements}
The authors would like to acknowledge Grant No. W911NF2220222 from the U.S. Army
Research Office, and the Center for Research Computing at the University
of Notre Dame. The authors report no conflict of interest.

\bibliographystyle{jfm}
\bibliography{zLibrary}

\end{document}